\documentclass[12pt]{article}
\newif\ifpdf            
\ifx\pdfoutput\undefined \pdffalse \else \pdftrue \fi
\ifpdf 
\usepackage[pdftex]{graphicx}
\pdfcompresslevel9 \DeclareGraphicsExtensions{.jpg,.pdf,.png,.mps}
\else 
\usepackage[dvips]{graphicx}
\DeclareGraphicsExtensions{.eps,.ps} \fi
\usepackage{bm,amssymb}

\title{Nematic elastomers with aligned carbon nanotubes: new electromechanical actuators}

\author
{S. Courty, J. Mine, A.R. Tajbakhsh, E.M. Terentjev$^\ast$\\
\\
\normalsize{Cavendish Laboratory, University of Cambridge,}\\
\normalsize{Cambridge CB3 0HE, U.K.}\\
\\
\normalsize{$^\ast$To whom correspondence should be addressed;
E-mail:  emt1000@cam.ac.uk} }
\date{}

\begin{document}
 \baselineskip24pt
 \maketitle

\begin{abstract}
  We demonstrate, for the first time, the large electromechanical
  response in nematic liquid crystalline elastomers filled with a
  very low ($\sim 0.01\%$) concentration of carbon nanotubes, aligned
  along the  nematic director at preparation. The nanotubes
  create a very large effective dielectric anisotropy of the
  composite. Their local field-induced torque is transmitted to
  the rubber-elastic network and is registered as the exerted
  uniaxial stress of order $\sim 1$~kPa in response to a constant
  field of order $\sim 1$~MV/m. We investigate the dependence of
  the effect on field strength, nanotube concentration and
  reproducibility under multiple field-on and -off cycles. The
  results indicate the potential of the nanotube-nematic elastomer
  composites as electrically driven actuators.
\end{abstract}

\paragraph*{Introduction}
The direct conversion of electrical energy to mechanical work
through a material response is critically important for a great
number of applications ranging from robotics and microscopic
pumps/valves, to biological muscle replacements. Over the years,
different kinds of actuators have been developed, with a very
broad range of characteristics of generated stress and stroke, as
well as the mechanism of actuation \cite{fleck97}. Liquid
crystalline elastomers \cite{WT03} have only recently entered the
actuator arena; with strains of up to 3-400\% and characteristic
stress of up to 1~MPa, nematic elastomers offer a range of new
engineering possibilities. In all cases studied so far, the
mechanism of actuation, represented as spontaneous uniaxial
extension/contraction of the nematic rubber along the director
axis, has been the coupling of the elastic network to the
underlying nematic order parameter $Q$. For prolate polymer
chains, with the radius of gyration along the director greater
than that perpendicular to it, $R_\| > R_\bot$, the relative
change in sample length with respect to the isotropic phase is
shown to be $L/L_0 = (R_\|/R_\bot)^{2/3} \approx (1+\alpha Q)$.
The last relation becomes a much sharper exponential dependence on
$Q$ in highly anisotropic main-chain nematic elastomers, see
\cite{WT03} for detail. Changing the order parameter $Q$ can be
achieved by heating the material, leading to thermal actuators
\cite{wermter,LofT,ratna}, by UV-light in photochromic materials
\cite{Finkphoto,Hogan:02}, or by other means such as dilution by
solvents.

Electrically-driven actuation in nematic elastomers has so far
been impossible, for a very simple reason. Assuming the electric
field $E$ interacts with dielectrically anisotropic nematic
medium, causing the director to rotate, the characteristic energy
density could be estimated as $\sim \varepsilon_{\rm o} \Delta
\varepsilon \, E^2$ \cite{deGennes93}. For a field $E\sim
10^6$~V/m and typical dielectric parameters of nematic liquid
crystals, this gives a density of $\sim 10^3~\hbox{J/m}^3$. The
rubbery elastic network resists any such rotation with a
characteristic energy density of the order of rubber modulus $\mu
\sim 10^5~\hbox{J/m}^3$ or higher; clearly, no effect could be
expected. In this paper we propose a new approach to this problem,
in essence, producing a composite material with an effective
dielectric anisotropy many orders of magnitude higher than in
usual liquid crystals. We achieve this by embedding a low
concentration of carbon nanotubes, aligned along the uniaxial
director of the monodomain nematic elastomer network. As a result,
we demonstrate a significant electromechanical response and study
its basic properties.

Since their discovery by S. Iijima in 1991 \cite{Iijima91}, the
carbon nanotubes (representing the fourth allotrope of carbon,
after diamond, graphite and fullerene) have been a subject of
great number of studies in all imaginable walks of science. Carbon
nanotubes (CNT) are formed by rolling up a graphite sheet into a
cylinder and capping each end with half of a fullerene molecule.
Depending on lattice coordination, CNT's are divided into several
structural and electronic classes. Not going into detail of this
field, very well covered in the literature \cite{cnt1,cnt2}, we
only need to emphasize the large mobility of $\pi$-electrons along
the cylinder axis, leading to an extremely high anisotropy of
polarizability. The mechanical strength and toughness of CNT's,
especially their multiwall variety, have been much celebrated in
the literature too, with the Young modulus of $\sim 1$~TPa and
withstanding the elastic bending by up to 70${}^{\rm o}$. CNT's
have been demonstrated to act as thermally, chemically or
electrically driven actuators on their own \cite{actu1,actu2},
individually or in macroscopic sheets and bundles \cite{actu3}. An
extremely interesting early report of mechanical actuation of a
CNT-polymer composite changing of temperature \cite{Koerner}
echoes much of our work reported here. However, we aim at the
electrical stimulation and explore the `internal degree of
freedom' end of this problem, using multiwall CNT's as passive
enhancers of natural dielectric anisotropy of a nematic rubber
shape-memory actuator.

The effective utilization of CNT's in polymer composite
applications depends strongly on the ability to homogeneously
disperse throughout a polymer matrix without destroying their
integrity. Many research groups have reported successes in this,
but as far as we could see, there is no robust and reliable method
of CNT dispersing. Another key problem is the alignment in the
polymer matrix. With the width of few nanometers and the length
approaching microns, the alignment of CNT's in a matrix is of
crucial importance. There are reports of alignment achieved by the
electric field \cite{field}, confirming our expectation of high
effective dielectric anisotropy. In our case the nanotubes were
dispersed at very low concentration in a gel network, which was
uniaxially stretched and then crosslinked further, to freeze in
the monodomain orientation of the nematic director. Naturally, the
CNT's were aligned in this direction as well. After this
preparation, the samples of nematic elastomer-CNT composite were
subjected to a constant and an oscillating electric field
perpendicular to the initial director/CNT alignment, Fig.~1. As a
result of strong dielectric torque on individual nanotubes, the
whole polymer network structure experienced a significant
mechanical shape change. In the fixed-length (isostrain)
conditions, we registered an actuator force of sample contraction
along the initial alignment axis.

\paragraph*{Materials and preparation}

The side-chain polysiloxane nematic polymers and their elastomer
networks, aligned in the uniaxial monodomain orientation, were
prepared in the Cavendish Laboratory following the procedure
pioneered by H. Finkelmann \cite{WT03,kupfer91}. The polysiloxane
backbone chains ($\sim 60$ monomer units long) had their Si$-$H
bonds reacted, using platinic acid catalyst, with the terminal
vinyl groups of the mesogenic rod-like molecule $4-$methoxyphenyl)
$-4'-$buteneoxy benzoate (MBB) and the two-functional crosslinker
$1$,$4-$di-$11-$undeceneoxy benzene (11UB), with the molar ratio
18:1 (thus achieving the 9:1 ratio of substituted groups on each
chain, or 10\% crosslinking density).

As has been mentioned above, the main difficulty is the separating
of CNT's from their solid aggregates and their dispersing in the
matrix at low concentration. To achieve this, we initially stirred
the nanotubes (kindly synthesized and supplied by Dr. C. Singh, of
Material Science, Cambridge) in toluene, using alternating
high-power sonication and mechanical stirrer at 2000~rpm for at
least 24 hours, before the addition of catalysts. The sample was
transferred into the polymer reacting mixture and stirred for a
further 4-6 hours. Finally, the crosslinking was initiated by a
combination of adding the catalyst and heating to 80${}^{\rm o}$C.
The subsequent procedure of two-stage crosslinking, with an
intermediate stretching to induce good director alignment, is well
described in the literature. For this pilot investigation, we
prepared three samples from the same batch, with identical
chemical composition and alignment, and with CNT concentration of
0, 0.0085 and 0.02 weight\%. At such low amount of loading, well
below the percolation limit \cite{Sandler}, one should expect each
nanotube to act on its own, embedded in a rubber elastic medium
and providing a very strong local anchoring to the nematic
director.

\paragraph*{Experimental details}

The materials were characterized with differential scanning
calorimetry and optical microscopy, confirming that no other phase
than the nematic were present below the clearing point $T_c
\approx 100^{\rm o}$C. The spontaneous uniaxial thermal expansion
measurements were performed with a travelling microscope focused
on a sample heated in the glass-fronted oven.

All mechanical measurements were performed at a fixed temperature
of $24^{\rm o}$C ($\pm 0.3^{\rm o}$), sufficiently far from the
glass, or nematic-isotropic transition points.  The dynamometer
setup is based on a custom-built device measuring the force
exerted by the sample in extension, at the controlled length and
temperature, with the accuracy $\pm$~4$\cdot 10^{-5}$~N
($\pm$~0.4~mg). The rectangular samples ($\sim
7\times$15$\times0.4$~mm) were mounted with insulating plastic
clamps, without any pre-stress, and not allowed to change shape
during the force-actuation experiments. Therefore, the exerted
stress was easily calculated as the measured force divided by the
fixed cross-section area $\sim 2.8~\hbox{mm}^2$. The electric
field was generated by a high power supply providing a voltage on
two aluminium plates separated by 1.56~mm and surrounding the flat
strip of clamped elastomer sample. With this setup, in order to
generate a substantial electric field, we needed a high voltage
(we used up to 3500~V) and so safety against discharges was
ensured by a number of additional electronic circuits.

\paragraph*{Sample structure}

The presence of CNT's in the rubbery network can be immediately
seen by the color of the otherwise transparent material. At
0.0085\% loading the elastomer is lightly grey and at 0.02\% the
material is dark grey and nearly opaque. Other samples, at much
higher nanotube loading, were completely black. We do not present
studies of these materials here because we could not verify the
quality of dispersing or the alignment of CNT's at higher
concentrations. Figure~2 shows the scanning electron microscopy
images of freeze-fractured composite samples, illustrating their
low concentration and good alignment at preparation.

It is interesting to see the role of low-concentration highly
anisotropic aligned filler on the nematic behavior of elastomer
network. Figure~3 shows ``classical'' thermal expansion
measurements (the thermal actuation of nematic rubbers, on very
slow cooling and heating), making comparison between the pure and
the composite samples. Clearly, the qualitative behavior is not
changed, with a constant sample length in the isotropic phase and
a rapid elongation in the nematic phase. This transition is
frequently reported in the literature as a diffuse increase, while
we see a nearly critical response: in fact, both curves in
Fig.~3(a) are fitted with the same law $L/L_0-1
=C(1-T/T_c)^{0.29}$. Surprisingly and unexpectedly to us, the
overall expansion/contraction magnitude (measured by the constant
$C$ in the above fit) was lower for the aligned CNT composite
elastomer. Since the nanotubes provide local sources of extra
alignment, we cannot attribute this to a reduction in nematic
order due to added impurities (in fact, their effect is probably
minor since the transition temperature has hardly changed). Our
explanation for the reduction in overall thermal expansion is
based on the assumption that nanotubes, which remain rigid and
aligned while embedded in the rubbery network, prevent the latter
from excessive shape changes along the alignment axis. Clearly,
this conclusion is only preliminary and the phenomenon has to be
investigated further, with varying concentration and length of
CNT's, as well as the natural capacity for expansion of the
network. One cannot exclude a possibility of rigid CNT's locally
rupturing the network, if its polymer chains are too highly
anisotropic; this was not the case in our studies because all
reported effects were reproducible after many cycles of
temperature, or later -- electric field.

The important characteristic of any rubber-elastic material is its
modulus. For a uniaxial nematic elastomer one may find it
difficult to identify a single value for the rubber modulus: the
matters of so-called ``soft elasticity'' form a central part of
their unique mechanical properties \cite{WT03}. However, for our
main experiments we need the particular geometry of uniaxial
extension along the axis of director (and CNT's) alignment.
Figure~4 shows that the linear stress-strain regime persists up to
the strain of at least 25\% and thus the linear (Young) modulus
can be unambiguously defined. Clearly the presence of aligned
CNT's makes the polymer composite stiffer. At the low
concentrations we are working at the Young modulus remains within
the same order of magnitude, $Y \sim 1$~MPa.

\paragraph*{Electromechanical response}

The samples, mounted on the insulating (thermally and
electrically) frame, which keeps the natural length of the sample
fixed but measures the contracting force exerted on the clamps,
were subjected to a constant electric field perpendicular to the
initial director (and CNT) orientation. With the applied voltage
of 3000~V, the field was $E\approx 1.9\cdot 10^6$V/m. Figure~5
shows the typical sequence of ``field-on'' and ``field-off''
cycles applied in the successive 1-hour sequence, demonstrating
the dependence of actuation stress on the elapsed time. The
response is clearly significant and, on the whole, follows the
expected trend. The on-cycle shows an immediate steep rise in
exerted stress, which then reaches a plateau level $\sigma_{\rm
max}$ that depends on CNT concentration as well as the field
strength; the latter is illustrated in Fig.~6.

We find no electromechanical response in the elastomer without
nanotubes, which confirms our estimates about the relative
strength of dielectric and elastic torques. With increasing CNT
concentration, the response becomes significant and reproducible.
In fact, we only show four cycles in this plot for clarity, while
the repeatability has been tested over many such cycles.

The mechanism of the electromechanical response seems to be clear:
the highly polarisable nanotubes experience a high torque to
rotate towards the direction of constant field, proportional to
$\Delta \varepsilon_{\rm CNT} E^2$. Nanotubes respond, and take
the elastic network with them, causing the measurable stress on
the clamped sample. It is possible that, at high enough field, the
nanotubes would rupture the local polymer network in their
vicinity. In our case the actual rotation of CNT's is small, which
is indicated by the field dependence of plateau stress
$\sigma_{\rm max}$, Fig.~6. If a full 90${}^{\rm o}$-rotation was
achieved one would see a plateau, whereas the linear increase
$\sigma_{\rm max} = {\rm const} \cdot E$ clearly indicates the
regime of ``small fields''.

The fact that the stress on the off-cycle does not quite return to
the initial zero level is an indication of how much we do not
understand about the underlying mechanism. This feature is
unambiguously reproduced for the 0.02\% sample, as the response at
each cycle reaches the same maximum level and falls down to the
same non-zero level when the field is off, thus confirming that no
sample degradation occurs. (The fact that we see no such effect
for the 0.0085\% sample could be simply due to the overall lower
amplitudes -- or possibly due to the cooperative effects of CNT
motion showing more prominently at their higher loading.) A
possible explanation is that, after spending an hour in the
strained state with the field on, in the $E$-off state the
viscoelastic polymer network has a very slow relaxation mode and
takes a much longer time to return to its initial equilibrium
(with zero stress).

An important feature of any actuator is the reproducibility and
the speed of its response. Figure~7 demonstrates the effect, for
the 0.02\% composite, under the field of $1.9\cdot 10^6$V/m (as in
Fig.~5) switching on and off every 30 seconds. Clearly the stress
responds very fast, to both ``on'' and ``off'' cycles. The initial
spike at each cycle is an artefact of our power supply electronics
that produces a voltage surge before settling at the required
constant value. Both the short-time and the long-time
representations in plots (a) and (b) illustrate a good
reproducibility of response speed and the plateau stress amplitude
$\sigma_{\rm max}$. This is practically useful for any actuator
application.

\paragraph*{Conclusions}

In this paper, for the first time, we have reported the
observation of significant electromechanical effect in liquid
crystal elastomer system filled with aligned carbon nanotubes. We
used the multiwall CNT's with no special purification, at very low
concentration (below percolation limit, so that samples remain
electrically insulating). The presumed mechanism of the response
is the very large anisotropy of polarizability of CNT's, which
results in the high local torque when an electric field
perpendicular to their equilibrium orientation is applied to the
sample. Small rotation of nanotubes is transmitted to the nematic
elastomer network -- even a small re-alignment of the director
induced by the CNT rotation generates a uniaxial mechanical
response of the whole sample, which we register as the actuation
stress.

The response stress is demonstrated to follow the linear
dependence on applied field, at least in the range of fields
studied. This, in fact, is puzzling because if one assumes the
angle of local CNT rotation, $\theta$, to be linearly proportional
to the dielectric torque (which in turn is $\propto \Delta
\varepsilon_{\rm CNT} E^2$), then the uniaxial strain induced in
the rubbery network along the initial director axis should be
$\propto (1-\cos \theta) \propto \theta^2$, which means the strain
(or in the isostrain conditions of our experiments, the exerted
stress $\sigma_{\rm max}$) could be expected to follow $\propto
E^4$; this is very clearly not the case.

Another difficulty we have is explaining the stress not returning
to zero after the field is switched off. This is most likely a
result of complex relaxation of polymer network with embedded
CNT's, after spending a long time (1 hour in Fig.~5) under the
applied field. However, simply claiming the slow relaxation as the
explanation of non-zero ``off''-stress may not be enough: if the
nanotubes start each following field-on cycle with a non-relaxed
pre-tilt angle, one might expect the ``on''-stress to increase.
The results of Fig.~5 clearly demonstrate that this is not the
case and that $\sigma_{\rm max}$ is a unique and reproducible
function of applied field, but not the field application history.
Note that in the fast-alternating field experiment on the same
material under the same field, Fig.~7, we see no residual stress
in the field-off state. If anything, the ``off''-stress is below
zero during the first hour of testing, perhaps due to a reactive
effect in the elastic network. Figure~7(b) shows that after the
initial settling period, the response becomes very regular,
returning to zero on every 30-second cycle.

We only studied composite materials with aligned CNT's at two low
concentrations. The main difficulty in preparation of samples with
higher loading was the CNT dispersing, which in our case was not
fully successful at higher concentrations. One may not be able to
exploit this electromechanical effect in this situation anyway
because above the percolation threshold (which is still very low,
$\sim 1$\% for highly anisotropic multiwall CNT's) the composite
would become a conductor.

Notwithstanding many shortcomings in our preparation techniques
and our understanding of the underlying fundamental physical
mechanisms, the effect of large electromechanical actuation is
demonstrated unambiguously and its speed and reproducibility make
it an attractive system for new applications. The uniaxial stress,
of the order $\sim 1$~kPa (or the corresponding strain of 0.1\%,
from the rubber modulus $\mu \sim 1$~MPa), is induced an electric
field of the order $\sim 1$~MV/m. There are many systems that
generate higher stress and the strains well above 50\% are common
in nematic elastomers. However, the ability to produce a
mechanical response by an electrical stimulus is invaluable for
many practical applications. Carbon nanotubes provide such a
capacity in liquid crystal elastomers. \\

This research was carried out during the Visiting Fellowship
(Stage de Magist\`ere) of JM, provided by {\sc Universit\'e de
Paris Sud -- Orsay}. We thank EPSRC (UK) and EOARD (US) for the
financial support. Help and advice of Hilmar Koerner and
Charanjeet Singh is gratefully appreciated. We particularly thank
Frank Baker for help with electron microscopy and Chau Nguen for
critical reading of the manuscript.


\clearpage

\begin{figure}
\centering
\resizebox{0.6\textwidth}{!}{\includegraphics{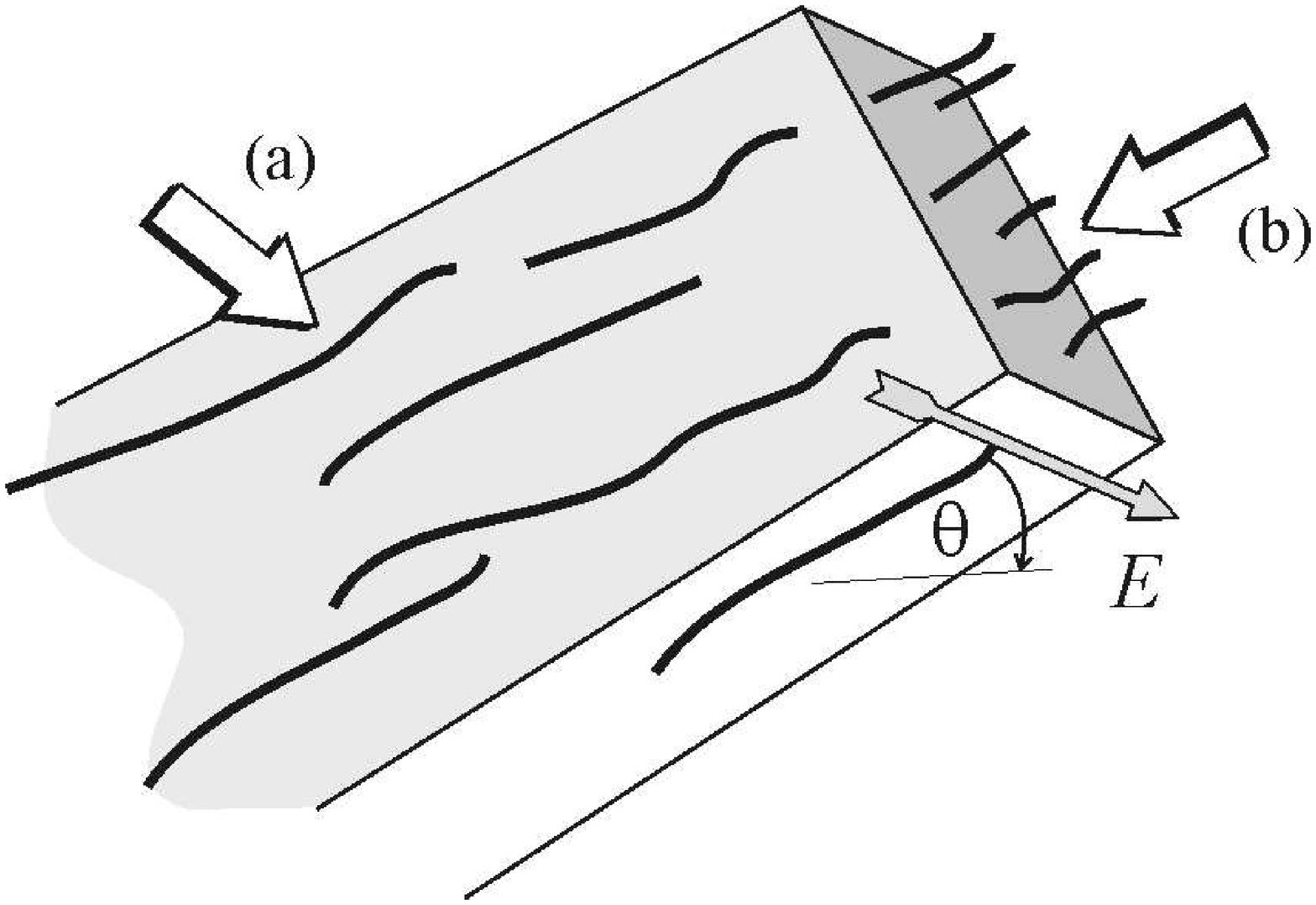}}
\caption[]{The sketch of a plane strip of nematic elastomer with
embedded aligned CNT's. Applying electric field $E$ across the
sample causes a local torque and a small nanotube rotation. Arrows
(a) and (b) indicate two planes of freeze-fracture, shown in
Fig.~2. } \label{sketch}
\end{figure}

\clearpage

\begin{figure}
\centering
\resizebox{0.85\textwidth}{!}{\includegraphics{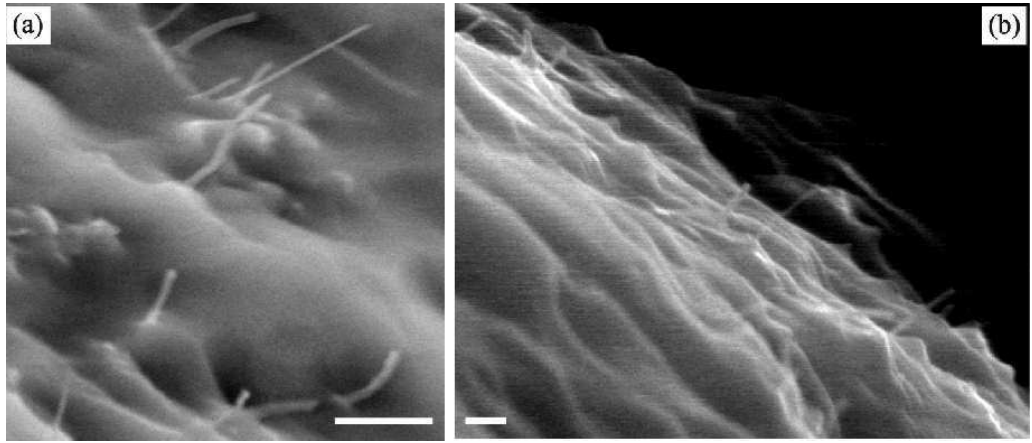}}
\caption[]{Electron microscopy images of 0.02\% CNT sample
freeze-fractured in different orientations, indicated in Fig.~1:
(a) in the plane parallel to the nematic director and the nanotube
alignment axis, (b) in the plane perpendicular to this axis. The
scale bar in both images is $0.5~\mu$m. One can see the CNT
alignment in (a). Image (b) shows several nanotubes protruding
perpendicularly out of the fracture surface, still keeping their
orientation.} \label{emic}
\end{figure}

\clearpage

\begin{figure}
\centering
\resizebox{0.99\textwidth}{!}{\includegraphics{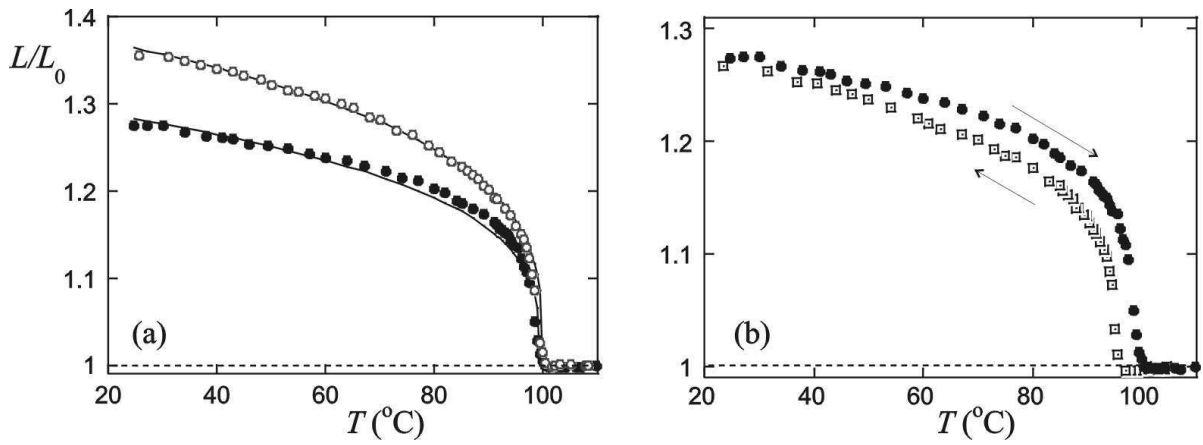}}
\caption[]{Uniaxial thermal expansion of nematic elastomers. (a)
The comparison, on heating, between the pure network ($\circ$) and
the 0.02\% CNT composite ($\bullet$). The solid lines are the
critical fit described in the text. (b) The hysteresis on heating
and cooling of the 0.02\% CNT composite. (No noticeable hysteresis
is recorded for the pure nematic elastomer.) } \label{thermal}
\end{figure}

\clearpage

\begin{figure}
\centering
\resizebox{0.6\textwidth}{!}{\includegraphics{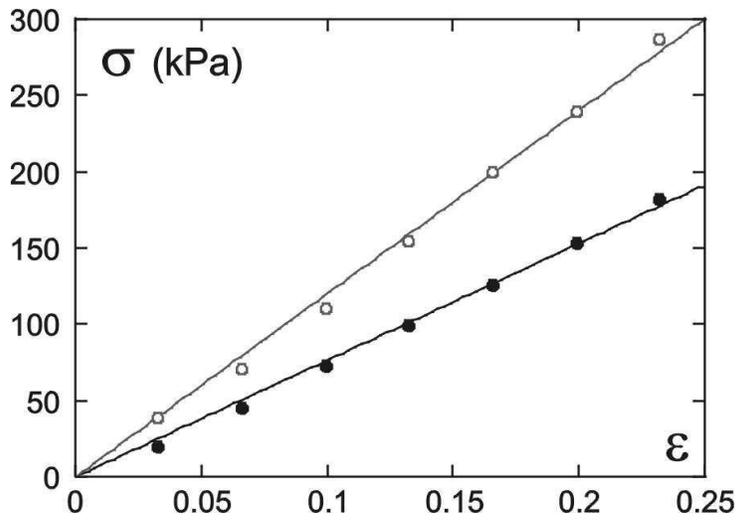}}
\caption[]{Equilibrium stress-strain variation on uniaxial
extension $\varepsilon = \Delta L/L$ along the director (and CNT
alignment) axis, for the pure network ($\circ$) and the 0.02\% CNT
composite ($\bullet$). The solid lines of linear fit provide the
Young (extension) modulus $Y_0 \approx 0.76$~MPa and $Y_{\rm CNT}
\approx 1.2$~MPa for the two systems.} \label{modulus}
\end{figure}

\clearpage

\begin{figure}  
\centering
\resizebox{0.6\textwidth}{!}{\includegraphics{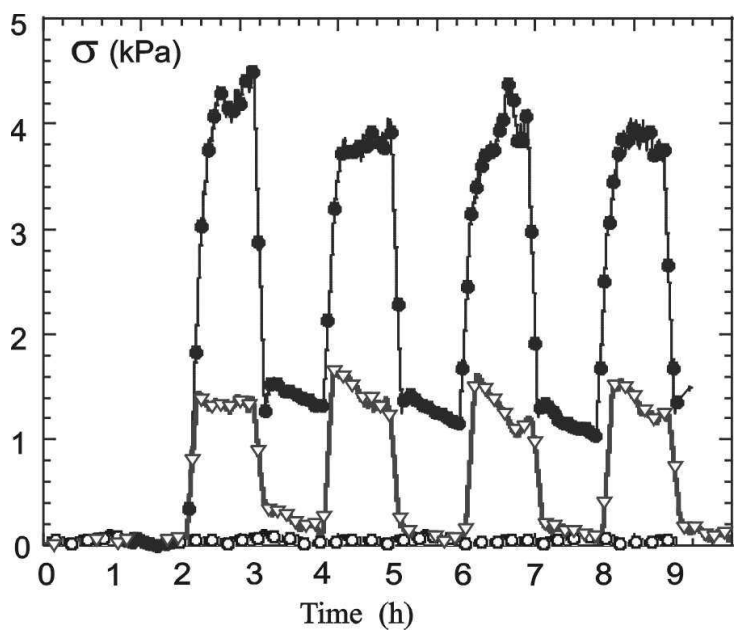}}
 \caption[]{A sequence of field cycles ($E\approx 1.9$~MV/m),
for samples with increasing CNT concentration: 0\% (non-filled
elastomer, $\circ$), 0.0085\% ($\triangledown$) and 0.02\%
($\bullet$). The plot indicates the role of increasing CNT loading
and also the equilibrium nature of the effect, reproducible after
many cycles. }   \label{conc}
\end{figure}

\clearpage

\begin{figure}  
\centering
\resizebox{0.6\textwidth}{!}{\includegraphics{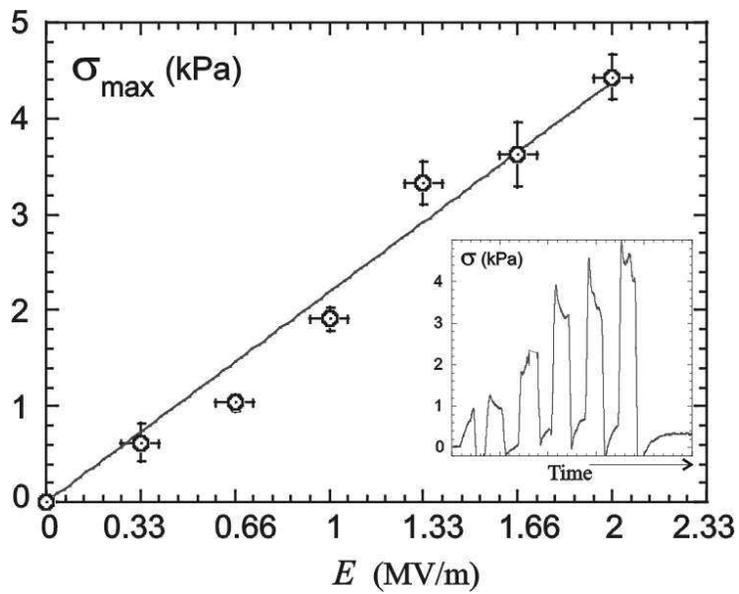}}
 \caption[]{Dependence of the plateau stress
 $\sigma_{\rm max}$ on the applied field
 strength, for the 0.02\% composite sample. The inset
 shows the elapsed time dependence, with each cycle at increasing constant
 field $E$. Solid lines show the fit $\sigma_{\rm max}(\hbox{kPa})=2.3 E(\hbox{MV/m})$. }
  \label{field}
\end{figure}

\clearpage

\begin{figure}  
\centering
\resizebox{0.99\textwidth}{!}{\includegraphics{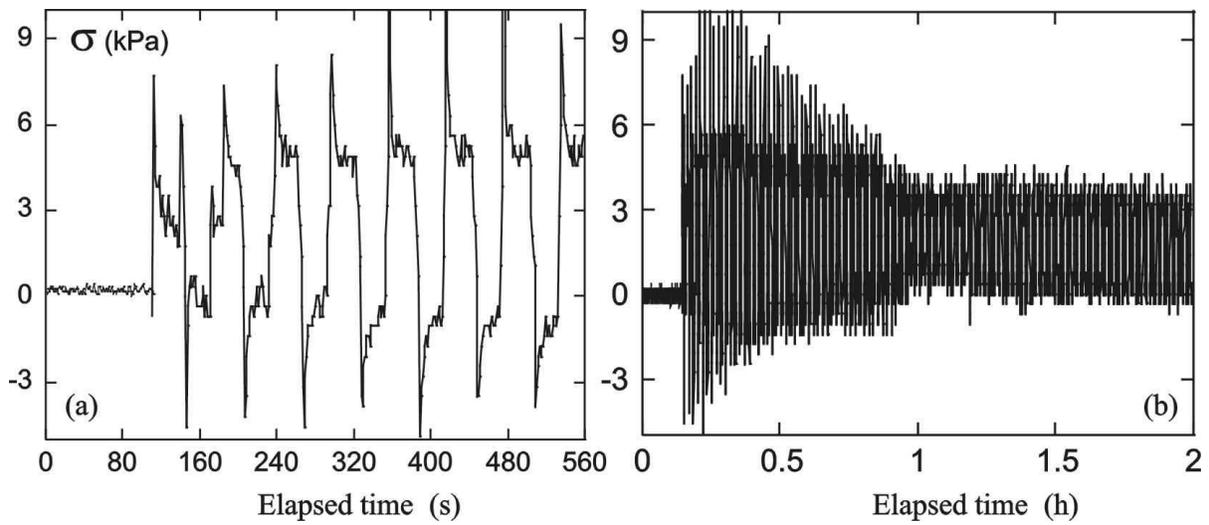}}
 \caption[]{The response of 0.02\% CNT composite to an alternating field,
 switching on and off every 30 seconds. The short initial period of this
 test, plot (a), illustrates the high speed of the stress response
 which closely follows the field spike produced by our power
 supply on both switching on and off. The long-time representation
 (b) cannot resolve each individual cycle, but demonstrates the
 overall stability of the response. } \label{oscil}
\end{figure}

\end{document}